\documentclass{article}
\usepackage{amsmath}
\usepackage[dvips]{graphics}
\usepackage{latexsym}
\usepackage{subfigure}
\DeclareFontFamily{OT1}{rsfs}{}
\DeclareFontShape{OT1}{rsfs}{m}{n}{ <-7> rsfs5 <7-10> rsfs7 <10->
; ; ; ; ; ; ; ; ; ; rsfs10}{}
\DeclareMathAlphabet{\mycal}{OT1}{rsfs}{m}{n}

\newcommand{\bea}{\begin{eqnarray*}}
\newcommand{\eea}{\end{eqnarray*}}
\newcommand{\bean}{\begin{eqnarray}}
\newcommand{\eean}{\end{eqnarray}}
\newcommand{\eqs}[1]{Eqs. (\ref{#1})}
\newcommand{\eq}[1]{Eq. (\ref{#1})}
\newcommand{\meq}[1]{(\ref{#1})}

\newcommand{\ppn}[2]{\frac{\partial #1}{\partial #2}}

\newcommand{\grad}{\nabla}

\newcommand{\eqn}{&=&}
\newcommand{\non}{\nonumber \\}

\newcommand{\sgt}{\sqrt{h}}
\newcommand{\hsp}{\hspace{0.1mm}}

\newcommand{\rt}{R^{(3)}}

\newcommand{\tem}{(T^{\textrm EM})}
\newcommand{\ta}{\tilde A}

\title{Proof of entropy principle in Einstein--Maxwell theory}
\author{Xiongjun Fang\thanks{Email: damiao\_ 2008@mail.bnu.edu.cn },
\ Sijie Gao\thanks{Corresponding author. Email: sijie@bnu.edu.cn}\\
Department of Physics, Beijing Normal University,\\
Beijing 100875, China}
\begin{document}
\maketitle

\begin{abstract}
We consider a static self-gravitating charged perfect fluid system in the Einstein--Maxwell theory. Assume Maxwell's equation and the Einstein constraint equation are satisfied and the temperature of the fluid obeys Tolman's law. Then, we prove that the extrema of total entropy implies other components of Einstein's equation for any variations of metric and electrical potential with fixed boundary values. Conversely, if Einstein's equation and Maxwell's equations hold, the total entropy achieves an extremum. Our work suggests that the maximum entropy principle is consistent with Einstein's equation when an electrostatic field is taken into account.

PACS number(s): 04.20.Cv, 04.20.Fy, 04.40.Nr

\end{abstract}

\section{Introduction}

It is well known that black holes can radiate and satisfy thermodynamical laws. This discovery establishes the connection between gravity and thermodynamics [1--6]. Despite the great success of black hole thermodynamics, there are still unresolved issues, for instance, the origin of black hole entropy. In contrast to black hole systems, local thermodynamic quantities of a perfect fluid in curved spacetimes, e.g., energy density $\rho$, entropy density $s$, and local temperature $T$, are well defined. The presence of gravity only affects the distribution of those local quantities.  In general there are two methods to determine the distribution of matter. One way is solving Einstein's equation. The other way is using the entropy principle to determine the distribution of matter.

Since entropy plays no role in Einstein's equation, it is unclear whether the two methods are consistent.  Even before the establishment of black hole thermodynamics, Cocke \cite{cocke} pointed out that the extrema of entropy should yield the equation of hydrostatic equilibrium which is derived from Einstein's equation. After that, Sorkin, Wald, and Zhang \cite{wald81} showed rigorously  that the Tolman--Oppenheimer--Volkoff equation of hydrostatic equilibrium can be derived from the extremum of total entropy and the Einstein constraint equation. Gao \cite{gao} extended their proof from radiation to a general perfect fluid, including uncharged fluid and uniformly charged fluid. This issue has been further explored in the past few years [10--18].

Recently, we \cite{fang} proved the entropy principle for a self-gravitating fluid in static spacetimes without any symmetry in the spacelike hypersurface. So far, the matter field considered is a perfect fluid. It is interesting to know whether or how the entropy principle works in the presence of an electromagnetic field. In this paper, we extend the two theorems in Ref. \cite{fang} to a uniformly charged fluid. The extension is not straightforward at all. For an uncharged perfect fluid, we have shown \cite{fang} that the variation of total entropy, $\delta S$,  is proportional to the variation of the spatial metric, $h_{ab}$. However, for a uniformly charged fluid, $\delta S$ appears to depend on $\delta h_{ab}$, $\delta\chi$, and $\delta A_a$, where $\chi$ is the redshift factor and $A^a$ is the vector potential of the electrostatic field. However, we manage to show that the components of $\delta\chi$ and $\delta h_{ab}$ vanish identically and the vanishing of the components of $\delta h_{ab} $ just gives the spatial components of Einstein's equation for a charged perfect fluid. Our work suggests that the entropy principle is consistent not only with the gravitational field but also with the electromagnetic field.

\section{Properties of charged perfect fluid in static spacetimes}

 We consider a general perfect fluid as discussed in Ref. \cite{gao}. The entropy density $s$ is taken to be a function of the energy density $\rho$ and particle number density $n$, i.e., $s=s(\rho,n)$. From the first law of thermodynamics, one can derive the integrated form of the Gibbs-Duhem relation,
\bean
s=\frac{1}{T}(\rho+p-\mu n)\,, \label{gb}
\eean
where $p$ and $\mu$ represent the pressure and the chemical potential, respectively. All the quantities are measured by static observers with 4-velocity $u^a$. These observers are orthogonal to the hypersurface $\Sigma$. Therefore, the induced metric on $\Sigma$ is given by
\bean
h_{ab}=g_{ab}+u_au_b\,.
\eean
The stress-energy tensor $T_{ab}$ for a perfect fluid takes the form
\bean
T_{ab}=\rho u_au_b+p h_{ab}=(\rho+p)u_au_b+p g_{ab} \,.   \label{sperfect}
\eean

Since the fluid is charged, we should also consider the stress energy tensor of the electromagnetic field,
\bean
\tem_{ab}=\frac{1}{4\pi}\left(F_{ac}F_b\hsp^c-\frac{1}{4}g_{ab}F_{de}F^{de}\right)\,, \label{tem}
\eean
where $F_{ab}=\grad_a\ta_b-\grad_b\ta_a$, and $\ta_a$ is the  vector potential. The electrostatic potential $\Phi$ is given by
\bean
\Phi=-\ta_a\xi^a=-\ta_a u^a\chi  \,,
\eean
where $\xi^a$ is the Killing vector and $\chi$ is the redshift factor.

The Maxwell's equation with source reads
\bean
\grad_b F^{ab} = 4\pi j^a = 4\pi \rho_e u^a\,,\label{maxeq}
\eean
where $j^a$ and $\rho_e$ represent the 4-current density of the electric charge and the charge density, respectively.
Then
\bean
 \xi^bF_{ab}\eqn \xi^b (\grad_a \ta_b-\grad_b \ta_a) \non
\eqn \grad_a(\xi^b\ta_b)-\ta_b\grad_a\xi^b-\xi^b\grad_b \ta_a \non
\eqn -\grad_a\Phi  \,,  \label{aphi}
\eean
where we have used the Killing equation and ${\cal L}_\xi \ta_a=0$ in the last step. From $\grad_{[a}F_{bc]}=0$ we can calculate
\bean
\grad_b \tem^{ab} \eqn  -j^c F^a\hsp_c \non
\eqn -\rho_e u^cF^a\hsp_c \non
\eqn -\frac{\rho_e}{\chi}\xi^cF^a\hsp_c \non
\eqn +\frac{\rho_e}{\chi}\grad^a\Phi\,,
\eean
where Maxwell's equation $\grad_{[a}F_{bc]}=0$ has been used in the first step and \eq{aphi} has been used in the last step.

We shall assume that Tolman's law holds, which states that the local temperature $T$ of the fluid satisfies
\bean
T\chi=T_0\,,  \label{tchi}
\eean
where $\chi$ is the redshift factor for static observers and $T_0$ is a constant. Without loss of generality,
we take $T_0=1$. This law establishes the relationship between the fluid temperature and the metric components.

It is then straightforward to show, from the conservation law $\grad_a[T^{ab}+\tem^{ab}]=0$ and the stationary conditions, that
\bean
0=\grad^ap+(\rho+p) A^a+\frac{\rho_e}{\chi}\grad^a\Phi  \,,
\eean
where $ A^a$ is the 4-acceleration of the observer. For stationary observers,
\bean
A_a=\grad_a\chi/\chi\,,
\eean
and thus
\bean
\grad_ap=-(\rho+p)\grad_a\chi/\chi-\frac{\rho_e}{\chi}\grad^a\Phi   \,. \label{gp1}
\eean

On the other hand, the local first law can be expressed in the form \cite{gao}
\bean
dp=sdT+nd\mu \,.  \label{gp2}
\eean
Using \eqs{tchi} and \meq{gb}, we find
\bean
\grad_ap\eqn \frac{\rho+p-\mu n}{T}\grad_aT+n\grad_a\mu \non
\eqn -\frac{\rho+p-\mu n}{\chi}\grad_a\chi+n\grad_a\mu\,.
\eean
Comparison with \eq{gp1} gives
\bean
-\frac{\rho_e}{n}\grad^a\Phi=\mu\grad_a\chi+\chi\grad_a\mu=\grad_a(\mu\chi)\,. \label{rhoemuchi}
\eean
If we assume that all particles possess the same charge $q$, i.e.,
\bean
\rho_e=n q\,,
\eean
then \eq{rhoemuchi} leads to
\bean
\mu\chi+q\Phi =c\,,
\eean
or
\bean
\frac{\mu}{T}+q\Phi =c\,,\label{mcqf}
\eean
where $c$ is a constant.

\section{Two theorems}
The distribution of a charged perfect fluid in static spacetimes can be determined in two ways. First, Einstein's equation and Maxwell's equation together can totally determine the distribution. Second, under certain boundary conditions, the total entropy of the fluid should take an extremum. In this section, we prove two theorems indicating the equivalence of the two methods.

\textbf{Theorem 1:}
Consider a uniformly charged perfect fluid in a static spacetime $(M,g_{ab})$ and $\Sigma$ as a three-dimensional hypersurface denoting a moment of the static observers. Let $C$ be a region on $\Sigma$ with a boundary $\bar C$. Let $h_{ab}$, $\Phi$, and $\chi$ be the induced metric on $\Sigma$, electrostatic potential, and  redshift factor, respectively. Assume that the temperature of the fluid obeys  Tolman's law and the Einstein constraint equation and  Maxwell's equation are satisfied in $C$. Then, the other components of Einstein's equation are implied by the extrema of the total fluid entropy for all variations of data in $C$ where $h_{ab}$, $\Phi$, $\chi$,  and their first derivatives are fixed on $\bar C$.

{\em Proof.}-- The total entropy $S$ is an integral of the entropy density $s$ over the region $C$ on $\Sigma$,
\bean
S=\int_C \sgt s(\rho,n)\,,
\eean
where $h$ is the determinant of $h_{ab}$.
 Thus, the variation of the total entropy is written in the form
\bean
\delta S=\int_C s\delta \sgt +\sgt\delta s\,.
\eean
Applying  the local first law of thermodynamics,
\bean
Tds=d\rho-\mu dn\,,
\eean
we find
\bean
\delta S\eqn \int_C s\delta \sgt +\sgt\left(\ppn{s}{\rho}\delta\rho+\ppn{s}{n}\delta n \right)\non
\eqn \int_C s\delta \sgt +\sgt\left(\frac{1}{T}\delta\rho-\frac{\mu}{T}\delta n \right) \label{dels}\,.
\eean
Note that $\mu/T$ is constant for an uncharged fluid and can be moved out of the integral \cite{fang}. But due to the electrostatic potential, \eq{mcqf} shows that $\mu/T$ is no longer a constant. We shall deal with the $\delta n$ term by employing Maxwell's equation.

Together with \eqs{gb} and \meq{mcqf}, we have
\bean
\label{dets}
\delta S=\int_C\frac{1}{T}(p+\rho-\mu n)\delta\sgt-\sgt(-q\Phi+c)\delta n+\sgt\frac{1}{T}\delta\rho \,.
\eean
Denote
\bean
\delta S=\int_C\delta L \,,
\eean
where
\bean
\delta L = \frac{1}{T}(p+\rho-\mu n)\delta\sgt-\sgt(-q\Phi+c)\delta n+\sgt\frac{1}{T}\delta\rho \,.\label{detL}
\eean

Our purpose is to derive the space components of Einstein's equation from $\delta L=0$ and the constraint Einstein equation. First, we need to express $\delta L$ as variations of basic variables $h_{ab}$, $\chi$ and $\Phi$.  The $\delta h$ term in \eq{detL} can be easily written in the desired form by the relation
\bean
\delta\sgt=\frac{1}{2}\sgt h^{ab}\delta h_{ab} \,.
\eean
The $\delta n$ term in \eq{detL} is calculated by using Maxwell's equation (see Appendix B). Now we shall focus on calculating the $\delta\rho$ term.

Note that the extrinsic curvature of $\Sigma$
defined by
\bean
\hat B_{ab}\equiv h^c_ah^d_b\grad_du_c
\eean
vanishes \cite{fang} in static spacetimes and
\bean
\label{uaab}
\grad_bu_a=-A_au_b\,.
\eean
By the result of Ref. \cite{fang},  the Ricci tensor $\rt_{ab}$ and scalar curvature $\rt$ of $\Sigma$ are given by
\bean
\rt_{ab}=R_{ab}+R_{aeb}\hsp^lu^eu_l+R_{fb}u^fu_a+R_{ak}u^ku_b+u_au_bR_{fk}u^fu^k
\eean
and
\bean
\rt=R+2R_{ab}u^au^b\,.
\eean

To calculate $\delta\rho$, we start with the Einstein constraint equation
\bean
G_{ab}u^au^b=8\pi T_{ab}^{total}u^au^b  \,, \label{ttotal}
\eean
where
\bean
T_{ab}^{total}=T_{ab}+T^{EM}_{ab} \,.
\eean
Together with \eqs{sperfect}, \meq{tem} and \meq{ttotal}, we find
\bean
\rho=\frac{1}{16\pi}\rt-\frac{1}{4\pi}(F_{ac}F_b\hsp^cu^au^b+\frac{1}{4}F_{ab}F^{ab}) \,. \label{rho}
\eean
Denote the last term of \eq{detL} by $\delta L_{\rho}$. By substituting \eq{rho} into \eq{detL}, we have
\bean
\delta L_{\rho}=\frac{1}{16\pi T}\sgt\delta\rt-\frac{1}{4\pi T}\sgt\left[\delta(F_{ac}F_b\hsp^cu^au^b)+\frac{1}{4}\delta(F_{ab}F^{ab})\right] \,.\label{detLrho}
\eean
By the standard calculation \cite{fang}, the first term on the right-hand side of \eq{detLrho} can be written in the desired form
\bean
\frac{1}{16\pi T}\sgt\delta\rt=\frac{\sgt}{T}(-\frac{1}{16\pi}\rt\hsp^{ab}+\frac{1}{16\pi}M_1^{ab})\delta h_{ab} \,, \label{detLrho1}
\eean
where
\bean
M_1^{ab}=A^aA^b+D^bA^a-h^{ab}\grad_cA^c \, \label{m1ab}
\eean
and $D_a$ is the derivative operator associated with $h_{ab}$.
Denote the second term on the right-hand side of \eq{detLrho} by  $\delta L_{\rho2}$, i.e.,
\bean
\delta L_{\rho2}=-\frac{1}{4\pi T}\sgt\left[\delta(F_{ac}F_b\hsp^cu^au^b)+\frac{1}{4}\delta(F_{ab}F^{ab})\right] \,.\label{delLrho22}
\eean
The calculation of \eq{delLrho22} is given in Appendix A. From \eqs{facfbc} and \meq{detff}, we have
\bean
\delta L_{\rho2} \eqn \frac{\sgt}{4\pi}\left[2D_c(u^bF_b\hsp^c)\delta\Phi+u^cF_c\hsp^b(D^a\Phi)\delta h_{ab}
+\frac{2D^c\Phi D_c\Phi}{\chi^2}\delta\chi\right] \non
&& -\frac{\sgt}{4\pi}A_aF^{ab}u_b\delta\Phi+\frac{\sgt}{4\pi}\grad_{a}(\frac{F^{ab}}{T})\frac{u_b}{\chi}\delta\Phi \non
&& +\frac{\sgt}{8\pi T}F^a\hsp_cF^{bc}\delta h_{ab}-\frac{\sgt D_c\Phi D^c\Phi}{4\pi\chi^2}\delta\chi  \non
\eqn \frac{\sgt}{4\pi}\left[2D_c(u^bF_b\hsp^c)-A_aF^{ab}u_b+\grad_{a}(\frac{F^{ab}}{T})\frac{u_b}{\chi}\right]\delta\Phi \non
&& +\frac{\sgt}{4\pi}\left[u^cF_c\hsp^b(D^a\Phi)+\frac{1}{2}F^a\hsp_cF^{bc}\right]\delta h_{ab}
+\frac{\sgt D_c\Phi D^c\Phi}{4\pi\chi^2}\delta\chi \,.\label{detLrho2}
\eean
The substitution of \eqs{detLrho1} and \meq{detLrho2} into \eq{detLrho} yields
\bean
\delta L_{\rho} \eqn \frac{\sgt}{T}\left[-\frac{1}{16\pi}\rt\hsp^{ab}+\frac{1}{16\pi}(A^aA^b+D^bA^a-h^{ab}\grad_cA^c)\right]\delta h_{ab} \non
&& +\frac{\sgt}{4\pi}\left[2D_c(u^bF_b\hsp^c)-A_aF^{ab}u_b+\grad_{a}(\frac{F^{ab}}{T})\frac{u_b}{\chi}\right]\delta\Phi \non
&& +\frac{\sgt}{4\pi}\left[u^cF_c\hsp^b(D^a\Phi)+\frac{1}{2}F^a\hsp_cF^{bc}\right]\delta h_{ab}
+\frac{\sgt D_c\Phi D^c\Phi}{4\pi\chi^2}\delta\chi \,.\label{dLrho}
\eean

The substitution of \eq{dLrho} and \eqs{nfab} and \meq{dsdc} in Appendix B into \eq{detL} yields
\bean
\delta L \eqn \frac{1}{T}(p+\rho-\mu n)\delta\sgt-\sgt(-q\Phi+c)\delta n+\sgt\frac{1}{T}\delta\rho  \non
\eqn \delta L_\Phi+\delta L_{\chi}+\delta L_h \,,
\eean
where
\bean
\delta L_\Phi \eqn -\frac{1}{4\pi}\sgt D^b(u^cF_{cb})\delta\Phi \non
&& +\frac{\sgt}{4\pi}\left[2D_c(u^bF_b\hsp^c)-A_aF^{ab}u_b+\grad_{a}(\frac{F^{ab}}{T})\frac{u_b}{\chi}\right]\delta\Phi  \,,
\eean
\bean
\delta L_{\chi} \eqn -\frac{\sgt D_c\Phi D^c\Phi}{4\pi\chi^2}\delta\chi
-\frac{\sgt}{4\pi}(D_b\Phi)(D^b\Phi)\chi^{-2}\delta\chi  \non
&& +\frac{\sgt}{4\pi}\frac{2D^c\Phi D_c\Phi}{\chi^2}\delta\chi\,,
\eean
\bean
\delta L_h \eqn \frac{\sgt}{T}\frac{p+\rho-\mu n}{2}h^{ab}\delta h_{ab}-\frac{1}{4\pi}\sgt(D^a\Phi)\chi^{-1}D^b\Phi\delta h_{ab}  \non
&& +\frac{1}{8\pi}\sgt h^{ab}D_d\left[(\Phi-\frac{c}{q})\chi^{-1}D^d\Phi\right]\delta h_{ab} \non
&& +\frac{\sgt}{T}\left[-\frac{1}{16\pi}\rt\hsp^{ab}+\frac{1}{16\pi}(A^aA^b+D^bA^a-h^{ab}\grad_cA^c)\right]\delta h_{ab} \non
&& +\frac{\sgt}{4\pi}\left[u^cF_c\hsp^b(D^a\Phi)+\frac{1}{2}F^a\hsp_cF^{bc}\right]\delta h_{ab} \,.  \label{Lh}
\eean
It is obvious that $\delta L_{\chi}=0$, which shows that the variation of the redshift factor $\chi$ has no contribution to $\delta S$. Now we show that $\delta L_\Phi$ also vanishes. We calculate
\bean
\delta L_\Phi \eqn -\frac{1}{4\pi}\sgt D^b(u^cF_{cb})\delta\Phi
+\frac{\sgt}{4\pi}\left[2D_c(u^bF_b\hsp^c)-A_aF^{ab}u_b+\grad_{a}\left(\frac{F^{ab}}{T}\right)\frac{u_b}{\chi}\right]\delta\Phi  \non
\eqn \frac{\sgt}{4\pi}\left[D_c(u^bF_b\hsp^c)-A_aF^{ab}u_b+\grad_{a}\left(\frac{F^{ab}}{T}\right)\frac{u_b}{\chi}\right]\delta\Phi  \non
\eqn \frac{\sgt}{4\pi}\left[h^a\hsp_c\grad_a(u^bF_b\hsp^c)-A_aF^{ab}u_b
+\grad_{a}\left(\frac{F^{ab}}{T}\frac{u_b}{\chi}\right)-\frac{F^{ab}}{T}\grad_a\left(\frac{u_b}{\chi}\right)\right]\delta\Phi \non
\eqn \frac{\sgt}{4\pi}\left[\grad_c(u^bF_b\hsp^c)+u^au_c\grad_a(u^bF_b\hsp^c)-A_aF^{ab}u_b
+\grad_{a}(F^{ab}u_b)-\frac{F^{ab}}{T}\grad_a\left(\frac{u_b}{\chi}\right)\right]\delta\Phi \,. \non  \label{LPhi}
\eean
Here, we have used $T\chi=1$. The last term of \eq{LPhi} vanishes because
\bean
&& -\frac{\sgt}{4\pi}\frac{F^{ab}}{T}\frac{(\grad_au_b)\chi-u_b\grad_a\chi}{\chi^2} \non
\eqn -\frac{\sgt}{4\pi}F^{ab}(-A_bu_a-A_au_b) \non
\eqn \frac{\sgt}{4\pi}F^{[ab]}2A_{(a}u_{b)} = 0  \,.
\eean
Note that
\bean
\grad_au_c=-u_a A_c \,,
\eean
and hence
\bean
\delta L_\Phi \eqn \frac{\sgt}{4\pi}\left[\grad_c(u^bF_b\hsp^c)+u^au_c\grad_a(u^bF_b\hsp^c)
-A_aF^{ab}u_b+\grad_{a}(F^{ab}u_b)\right]\delta\Phi \non
\eqn \frac{\sgt}{4\pi}\left[u^au_c\grad_a(u^bF_b\hsp^c)
-A_aF^{ab}u_b\right]\delta\Phi \non
\eqn \frac{\sgt}{4\pi}\left[u^au_cF_b\hsp^c\grad_au^b
-A_aF^{ab}u_b\right]\delta\Phi \non
\eqn \frac{\sgt}{4\pi}\left[-u^au_cF_b\hsp^cu_aA^b
-A_aF^{ab}u_b\right]\delta\Phi \non
\eqn 0 \,.
\eean
This result reveals that the variation of the electrostatic
potential $\Phi$ has no contribution to $\delta S$. So we have
\bean
\delta L \eqn \delta L_h  \non
\eqn \frac{\sgt}{T}\frac{p+\rho-\mu n}{2}h^{ab}\delta h_{ab}-\frac{1}{4\pi}\sgt(D^a\Phi)\chi^{-1}D^b\Phi\delta h_{ab}  \non
&& +\frac{1}{8\pi}\sgt h^{ab}D_d\left[\left(\Phi-\frac{c}{q}\right)\chi^{-1}D^d\Phi\right]\delta h_{ab} \non
&& +\frac{\sgt}{T}\left[-\frac{1}{16\pi}\rt\hsp^{ab}+\frac{1}{16\pi}(A^aA^b+D^bA^a-h^{ab}\grad_cA^c)\right]\delta h_{ab} \non
&& +\frac{\sgt}{4\pi}\left[u^cF_c\hsp^b(D^a\Phi)+\frac{1}{2}F^a\hsp_cF^{bc}\right]\delta h_{ab} \,.
\eean
This shows explicitly that $\delta S$ is determined by the variation of $h_{ab}$ only. Since $\delta S=0$ by the assumption of Theorem 1,we have
\bean
&& \frac{\sgt}{T}\frac{p+\rho-\mu n}{2}h^{ab}-\frac{\sgt}{4\pi}(D^a\Phi)\chi^{-1}D^b\Phi  +\frac{\sgt}{8\pi}h^{ab}D_d\left[\left(\Phi-\frac{c}{q}\right)\chi^{-1}D^d\Phi\right] \non
&& +\frac{\sgt}{T}\left[-\frac{1}{16\pi}\rt\hsp^{ab}+\frac{1}{16\pi}(A^aA^b+D^bA^a-h^{ab}\grad_cA^c)\right] \non
&& +\frac{\sgt}{4\pi}\left[u^cF_c\hsp^b(D^a\Phi)+\frac{1}{2}F^a\hsp_cF^{bc}\right] =0  \,. \label{detseqn0}
\eean
By substituting \eq{rho} and \eqs{nfab} and \meq{muT} in Appendix B into \eq{detseqn0}, and letting $c'=-c/q$, we have
\bean
8\pi ph^{ab} \eqn -\frac{1}{2}\rt h^{ab}+2(F_{dc}F_e\hsp^cu^du^e+\frac{1}{4}F_{cd}F^{cd})h^{ab} \non
&& +2T(\Phi+c')h^{ab}u_c\grad_dF^{cd}-2Th^{ab}D_c[(\Phi+c')u_dF^{dc}] \non
&& +4T(\chi^{-1}D^b\Phi)D^a\Phi+\rt\hsp^{ab}-(A^aA^b+D^bA^a-h^{ab}\grad_cA^c)  \non
&& -4Tu^cF_c\hsp^bD^a\Phi-2h^{ac}h^{bd}F_{ce}F_d\hsp^e \,.\label{phab}
\eean

From Ref. \cite{fang} we already know that
\bean
h^{ab}R_{cd}u^cu^d-h^{ac}h^{bd}R_{ced}\hsp^lu^eu_l+A^aA^b+D^bA^a-h^{ab}\grad_cA^c=0 \,.\label{nocharge}
\eean
Substituting \eq{nocharge} into \eq{phab}, we have
\bean
8\pi ph^{ab} \eqn h^{ac}h^{bd}R_{cd}-\frac{1}{2}Rh^{ab}-2(F_{ce}F_d\hsp^eh^{ac}h^{bd}-\frac{1}{4}h^{ab}F_{cd}F^{cd}) \non
&& +P^{ab}_1+P^{ab}_2   \,,  \label{p}
\eean
where
\bean
P^{ab}_1\eqn 2F_{dc}F_e\hsp^cu^du^eh^{ab}+2T(\Phi+c')h^{ab}u_c\grad_dF^{cd}\non
&&-2Th^{ab}D_c[(\Phi+c')u_dF^{dc}]\,, \label{p1} \\
P^{ab}_2\eqn 4T(\chi^{-1}D^b\Phi)D^a\Phi-4Tu^cF_c\hsp^bD^a\Phi \,.  \label{p2}
\eean

Now, we show that $P^{ab}_1$ and $P^{ab}_2$ vanish. We first calculate
\bean
D_c(u_dF^{dc}) \eqn h_c\hsp^e\grad_e(u_dF^{dc}) \non
\eqn u_cu^e\grad_e(u_dF^{dc})+\grad_c(u_dF^{dc}) \non
\eqn u_cu^eF^{dc}\grad_eu_d+F^{dc}\grad_cu_d+u_d\grad_cF^{dc} \non
\eqn -u_cu^eF^{dc}u_eA_d-F^{dc}u_cA_d+u_d\grad_cF^{dc} \non
\eqn u_d\grad_cF^{dc} \,.
\eean
With the help of \eqs{aphi} and \meq{tchi}, \eq{p1} can be written as
\bean
P^{ab}_1 \eqn h^{ab}[2T^2\grad^c\Phi\grad_c\Phi+2T(\Phi+c')u_c\grad_dF^{cd}  \non
&& -2T(\Phi+c')D_c(u_dF^{dc})-2Tu_dF^{dc}D_c(\Phi+c')] \non
\eqn 2h^{ab}[T^2\grad^c\Phi\grad_c\Phi+T(\Phi+c')u_c\grad_dF^{cd}  \non
&& -T(\Phi+c')u_d\grad_cF^{dc}-T^2\grad^c\Phi\grad_c\Phi] \non
\eqn 0 \,.
\eean
Applying \eqs{aphi} and \meq{tchi} again, we find immediately
\bean
P^{ab}_2 =0 \,.
\eean
Therefore, \eq{p} just gives the projection of Einstein's equation on $\Sigma$
\bean
8\pi ph^{ab}=R_{cd}h^{ac}h^{bd}-\frac{1}{2}Rh^{ab}-2(F_{ce}F_d\hsp^e-\frac{1}{4}g_{cd}F_{ef}F^{ef})h^{ac}h^{bd}\,. \label{p}
\eean
This completes the proof of Theorem 1.

In the above proof, we used the Einstein constraint \eq{rho} to derive \eq{detL}. Then, by applying $\delta S=0$, we obtained the spatial components of Einstein's equation. It is not difficult to check that the proof is reversible; i.e., from the projected Einstein equation (64), one can show $\delta L=0$ in \eq{detL}, which makes the total entropy an extremum. Thus, we arrive at the following theorem.

\textbf{Theorem 2:}
Consider a perfect fluid with charge in a static spacetime $(M,g_{ab})$ and $\Sigma$ as a three-dimensional hypersurface denoting a moment of the static observers. Let $C$ be a region on $\Sigma$ with a boundary $\bar C$, $h_{ab}$, $\Phi$, and $\chi$ be the induced metric, potential, and redshift on $\Sigma$. Assume that the temperature of the fluid obeys Tolman's law and both Einstein's equation and Maxwell's equation are satisfied in $C$. Then, the fluid is distributed such that its total entropy in $C$ is an extremum for all variations where $h_{ab}$, $\Phi$, $\chi$, and their first derivatives are fixed on $\bar C$.

\section{Conclusions}
We have rigorously proven the equivalence of the extrema of the entropy and Einstein's equation under a few natural and necessary conditions. Different from the proof for the uncharged case, we have to consider the variations of $\chi$ and $A^a$. The treatment of the $\delta n$ term also totally differs from the uncharged case because $\mu/T$ is no longer a constant.  The significant improvement from previous works is that we extended the maximum entropy principle to the Einstein-Maxwell theory. Our work suggests a clear connection between Einstein's equation and the thermodynamics of a charged perfect fluid in static spacetimes.

\section*{Acknowledgements}
 This research was supported by NSFC Grants No. 11235003, No. 11375026 and No. NCET-12-0054.

\begin{appendix}

\section{Calculation of \eq{delLrho22}}
In this Appendix, we will show the detailed calculation of $-\frac{\sgt}{4\pi T}\delta(F_{ac}F_b\hsp^cu^au^b)$ and
$-\frac{\sgt}{16\pi T}\delta(F_{ab}F^{ab})$ in \eq{delLrho22}. First, we calculate
\bean
&& -\frac{\sgt}{4\pi T}\delta(F_{ac}F_b\hsp^cu^au^b)  \non
\eqn -\frac{\sgt}{4\pi T}\left[u^bF_b\hsp^c\delta(F_{ac}u^a)+u^aF_{ac}\delta(F_b\hsp^cu^b)\right] \non
\eqn -\frac{\sgt}{4\pi T}\left[u^bF_b\hsp^c\delta(\chi^{-1}D_c\Phi)+u^aF_{ac}\delta(\chi^{-1}D^c\Phi)\right]  \non
\eqn -\frac{\sgt}{4\pi T}\left[\chi^{-1}u^bF_b\hsp^c\delta(D_c\Phi)+u^bF_b\hsp^cD_c\Phi\delta\chi^{-1}
+\chi^{-1}u^aF_{ac}\delta(h^{dc}D_d\Phi)+u^aF_{ac}D^c\Phi\delta\chi^{-1}\right]\non
\eqn -\frac{\sgt}{4\pi T}\left[\chi^{-1}u^bF_b\hsp^cD_c\delta\Phi-\frac{2D^c\Phi D_c\Phi}{\chi^3}\delta\chi
+\chi^{-1}u^aF_{ac}h^{dc}D_d\delta\Phi+\chi^{-1}u^aF_{ac}(D_d\Phi)\delta h^{dc}\right]  \non
\eqn -\frac{\sgt}{4\pi}\left[u^bF_b\hsp^cD_c\delta\Phi-\frac{2D^c\Phi D_c\Phi}{\chi^2}\delta\chi +u^aF_{ac}h^{dc}D_d\delta\Phi+u^aF_{ac}(D_d\Phi)\delta h^{dc}\right]  \,,
\eean
where \eqs{aphi} and \meq{tchi} have been used.
Using integration by parts and dropping the boundary terms, we have
\bean
&& -\frac{\sgt}{4\pi T}\delta(F_{ac}F_b\hsp^cu^au^b) \non
\eqn \frac{\sgt}{4\pi}\left[D_c(u^bF_b\hsp^c)\delta\Phi+\frac{2D^c\Phi D_c\Phi}{\chi^2}\delta\chi
+D_d(u^aF_{ac}h^{dc})\delta\Phi+u^aF_a\hsp^c(D^d\Phi)\delta h_{dc}\right]  \non
\eqn \frac{\sgt}{4\pi}\left[2D_c(u^bF_b\hsp^c)\delta\Phi+\frac{2D^c\Phi D_c\Phi}{\chi^2}\delta\chi
+u^cF_c\hsp^b(D^a\Phi)\delta h_{ab}\right]  \,. \label{facfbc}
\eean
\\

Now we turn to $-\frac{\sgt}{16\pi T}\delta(F_{ab}F^{ab})$.
\bean
&&  -\frac{\sgt}{16\pi T}\delta(F_{ab}F^{ab})  \non
\eqn -\frac{\sgt}{16\pi T}(F^{ab}\delta F_{ab}+F_{ab}\delta F^{ab})  \non
\eqn -\frac{\sgt}{16\pi T}\left[F^{ab}\delta F_{ab}+F_{ab}\delta(F^{cd}g^{ac}g^{bd})\right]  \non
\eqn -\frac{\sgt}{16\pi T}\left[F^{ab}\delta F_{ab}+F_{ab}g^{ac}g^{bd}\delta F^{cd}
+F_{ab}g^{ac}\delta(F^{cd}g^{bd})+F_{ab}g^{bd}\delta(F^{cd}g^{ac})\right]  \non
\eqn -\frac{\sgt}{16\pi T}(2F^{ab}\delta F_{ab}+2F_{ac}F_b\hsp^c\delta g^{ab})  \,.
\eean
Because $\delta F_{ab}=2\grad_{[a}\delta\ta_{b]}$ and $g^{ab}=h^{ab}-u^au^b$, we have
\bean
&& -\frac{\sgt}{16\pi T}\delta(F_{ab}F^{ab}) \non
\eqn  -\frac{\sgt}{16\pi T}\left[4F^{ab}\grad_{[a}\delta\ta_{b]}+2F_{ac}F_b\hsp^c\delta h^{ab}
-2F_{ac}F_b\hsp^c\delta(u^au^b)\right] \non
\eqn  -\frac{\sgt}{16\pi T}\left[4F^{ab}\grad_{a}\delta\ta_{b}+2F_{ac}F_b\hsp^c\delta h^{ab}
-2F_{ac}F_b\hsp^cu^a\delta u^b-2F_{ac}F_b\hsp^cu^b\delta u^a\right]  \non
\eqn  -\frac{\sgt}{16\pi T}\left[4F^{ab}\grad_{a}\delta\ta_{b}+2F_{ac}F_b\hsp^c\delta h^{ab}
-2F_{ac}F_b\hsp^cu^a\xi^b\delta\chi^{-1}-2F_{ac}F_b\hsp^cu^b\xi^a\delta\chi^{-1}\right]  \non
\eqn -\frac{\sgt}{16\pi T}\left[4F^{ab}\grad_{a}\delta\ta_{b}+2F_{ac}F_b\hsp^c\delta h^{ab}
+\frac{4D_c\Phi D^c\Phi}{\chi^3}\delta\chi\right] \,.  \label{fabfab}
\eean
Since $\ta_a=-\Phi(dt)_a$, we have\footnote{ The coordinates are fixed for variations. So $\delta(dt)_a=0$}
\bean
\delta\ta_a \eqn -(dt)_a\delta\Phi=\frac{u_a}{\chi}\delta\Phi \,.
\eean
Using integration by parts for the first term in \eq{fabfab}, we have
\bean
&& -\frac{\sgt}{4\pi T}F^{ab}\grad_{a}\delta\ta_{b}  \non
\eqn -\frac{\sgt}{4\pi T}F^{ab}\grad_{a}\left(\frac{u_b}{\chi}\delta\Phi\right)\non
\eqn -\frac{\sgt}{4\pi}\grad_{a}\left(\frac{F^{ab}}{T}\frac{u_b}{\chi}\delta\Phi\right)
+\frac{\sgt}{4\pi}\grad_{a}\left(\frac{F^{ab}}{T}\right)\frac{u_b}{\chi}\delta\Phi \non
\eqn -\frac{\sgt}{4\pi}\grad_{a}\left(F^{ab}u_b\delta\Phi\right)
+\frac{\sgt}{4\pi}\grad_{a}\left(\frac{F^{ab}}{T}\right)\frac{u_b}{\chi}\delta\Phi \,.  \label{detphi}
\eean

Note that for any $v^a$ tangent to $\Sigma$, i.e., $v^a u_a=0$. Thus,
\bean
D_av^a \eqn h_a\hsp^ch_d\hsp^a\grad_cv^d \non
\eqn \grad_av^a-v^aA_a  \,,
\eean
or
\bean
\grad_av^a=D_av^a+v^aA_a \,.\label{DvvA}
\eean
Therefore,
\bean
&&-\frac{\sgt}{4\pi}\grad_{a}(F^{ab}u_b\delta\Phi) \non
\eqn -\frac{\sgt}{4\pi}D_{a}(F^{ab}u_b\delta\Phi)
-\frac{\sgt}{4\pi}A_aF^{ab}u_b\delta\Phi  \non
\eqn -\frac{\sgt}{4\pi}A_aF^{ab}u_b\delta\Phi\,,
\eean
where we have  dropped the boundary term in the last step.
Then \eq{detphi} becomes
\bean
&&  -\frac{\sgt}{4\pi T}F^{ab}\grad_{[a}\delta\ta_{b]}  \non
\eqn  -\frac{\sgt}{4\pi}A_aF^{ab}u_b\delta\Phi+\frac{\sgt}{4\pi}\grad_{a}\left(\frac{F^{ab}}{T}\right)\frac{u_b}{\chi}\delta\Phi  \,.
\eean
The substitution of this result into \eq{fabfab} gives
\bean
&& -\frac{\sgt}{16\pi T}\delta(F_{ab}F^{ab}) \non
\eqn  -\frac{\sgt}{16\pi T}\left[4F^{ab}\grad_{a}\delta\ta_{b}+2F_{ac}F_b\hsp^ch^{ab}
+\frac{4D_c\Phi D^c\Phi}{\chi^3}\delta\chi\right]  \non
\eqn  -\frac{\sgt}{4\pi}A_aF^{ab}u_b\delta\Phi+\frac{\sgt}{4\pi}\grad_{a}\left(\frac{F^{ab}}{T}\right)\frac{u_b}{\chi}\delta\Phi \non
&& +\frac{\sgt}{8\pi T}F^a\hsp_cF^{bc}\delta h_{ab}-\frac{D_c\Phi D^c\Phi}{4\pi\chi^2}\delta\chi \,.\label{detff}
\eean

\section{Calculation of the $\delta n$ term in \eq{detL}}
In this Appendix, we will show the detailed calculation of $-\sgt(-q\Phi+c)\delta n$. Define
\bean
\delta S_n = -\int_C\sgt(-q\Phi+c)\delta n = \int_C\delta L_n \,.
\eean
To express $\delta n$ as a combination of $\delta\chi$, $\delta\Phi$, and $\delta h_{ab}$, we need to employ Maxwell's equation. Suppose that all particles possess the same charge $q$,
\bean
\rho_e=q n \,.
\eean
From Maxwell's equation \eq{maxeq}, we have
\bean
n \eqn -\frac{1}{4\pi q}u_a\grad_b(F^{ab})  \,. \label{n}
\eean
Therefore, with the help of \eq{DvvA}
\bean
n \eqn -\frac{1}{4\pi q}u_a\grad_bF^{ab} \non
\eqn -\frac{1}{4\pi q}\grad_b(u_aF^{ab})+\frac{1}{4\pi q}F^{ab}\grad_b u_a \non
\eqn -\frac{1}{4\pi q}D_b(u_aF^{ab})-\frac{1}{4\pi q}u_aF^{ab}A_b-\frac{1}{4\pi q}F^{ab}u_b A_a \non
\eqn -\frac{1}{4\pi q}D_b(\chi^{-1}D^b\Phi)  \,.\label{nfab}
\eean

Since
\bean
\frac{\mu}{T}=-q\Phi+c  \label{muT} \,,
\eean\
where c is constant, we have
\bean
\delta L_n \eqn -\sgt(-q\Phi+c)\delta n \non
\eqn \frac{\sgt}{4\pi q}(-q\Phi+c)\delta[D_b(\chi^{-1}D^b\Phi)] \,.
\eean
Note
\bean
D_b(\chi^{-1}D^b\Phi)(\lambda)
\eqn D_b(0)(\chi^{-1}h^{bc}D_c\Phi)+C^b\hsp_{bd}(\lambda)\chi^{-1}D^d\Phi \,,
\eean
and then
\bean
\delta L_n \eqn \frac{\sgt}{4\pi q}(-q\Phi+c)\delta[D_b(\chi^{-1}D^b\Phi)] \non
\eqn \frac{\sgt}{4\pi q}(-q\Phi+c)D_b(D^b\Phi\delta\chi^{-1})
+ \frac{\sgt}{4\pi q}(-q\Phi+c)D_b(\chi^{-1}D^b\delta\Phi) \non
&&+ \frac{\sgt}{4\pi q}(-q\Phi+c)D_b(\chi^{-1}D_c\Phi\delta h^{bc})
+ \frac{\sgt}{4\pi q}(-q\Phi+c)(\chi^{-1}D^d\Phi)\delta C^b\hsp_{bd} \non
\eqn \frac{\sgt}{4\pi q}D_b[(-q\Phi+c)(D^b\Phi\delta\chi^{-1})]
-\frac{\sgt}{4\pi q}D_b(-q\Phi+c)(D^b\Phi\delta\chi^{-1})\non
&&-\frac{\sgt}{4\pi q}D_b(-q\Phi+c)\chi^{-1}D^b\delta\Phi
-\frac{\sgt}{4\pi q}D_b(-q\Phi+c)\chi^{-1}D_c\Phi\delta h^{bc}  \non
&& +\frac{\sgt}{4\pi q}(-q\Phi+c)(\chi^{-1}D^d\Phi)\delta C^b\hsp_{bd} \non
\eqn -\frac{\sgt}{4\pi}(D_b\Phi)(D^b\Phi)\chi^{-2}\delta\chi
-\frac{\sgt}{4\pi}D^b(\chi^{-1}D_b\Phi)\delta\Phi \non
&& +\frac{\sgt}{4\pi}(D_b\Phi)\chi^{-1}D_c\Phi\delta h^{bc}
-\frac{\sgt}{4\pi}(\Phi-\frac{c}{q})(\chi^{-1}D^d\Phi)\delta C^b\hsp_{bd}  \,,
\eean
where we have used integration by parts twice and discarded the boundary terms.
Since \cite{waldbook}
\bean
\delta C^b\hsp_{bd}=\frac{1}{2}h^{bc}D_d\delta h_{bc} \,,
\eean
we have
\bean
\delta L_n \eqn -\frac{\sgt}{4\pi}(D_b\Phi)(D^b\Phi)\chi^{-2}\delta\chi
-\frac{\sgt}{4\pi}D^b(\chi^{-1}D_b\Phi)\delta\Phi  \non
&& +\frac{\sgt}{4\pi}(D_b\Phi)\chi^{-1}D^b\Phi\delta h^{bc}  \non
&& -\frac{\sgt}{8\pi}h^{bc}(\Phi-\frac{c}{q})(\chi^{-1}D^d\Phi)D_d\delta h_{bc} \non
\eqn -\frac{\sgt}{4\pi}(D_b\Phi)(D^b\Phi)\chi^{-2}\delta\chi
-\frac{\sgt}{4\pi}D^b(u^cF_{cb})\delta\Phi  \non
&& -\frac{\sgt}{4\pi}(D^a\Phi)\chi^{-1}D^b\Phi\delta h_{ab}  \non
&& +\frac{\sgt}{8\pi}h^{ab}D_d\left[(\Phi-\frac{c}{q})\chi^{-1}D^d\Phi\right]\delta h_{ab} \,.  \label{dsdc}
\eean
Hence, $\delta L_n$ has been expressed as the linear combination of $\delta\chi$, $\delta\Phi$ and $\delta h_{ab}$.

\end{appendix}


\begin{thebibliography}{3}
\bibitem{b1} J. D. Bekenstein, Phys. Rev. D, {\bf 7}, 2333 (1973).
\bibitem{bardeen1973} J. M. Bardeen, B. Carter, and S.W. Hawking, Commun. Math. Phys. {\bf 31}, 161 (1973).
\bibitem{hawking75} S. W. Hawking, Commun.Math. Phys. {\bf 43}, 199(1975).

\bibitem{iyer94} V. Iyer and R. M. Wald, Phys. Rev. {\bf 50}, 846 (1994).
\bibitem{waldreview} R. M. Wald, Living Rev. Relativity {\bf 4}, 6(2001).
\bibitem{ted} T. Jacobson, Phys. Rev. Lett. {\bf 75},   1260 (1995).

\bibitem{cocke} W. J. Cocke, Ann. Inst. Henri Poincar$\acute{e}$ {\bf 2}, 283 (1965).
\bibitem{wald81} R. D. Sorkin, R. M. Wald and Z. J. Zhang, General Relativ. Gravit. {\bf 13}, 1127 (1981).
\bibitem{gao} S. Gao, Phys.Rev.D {\bf 84,} 104023(2011); {\bf 85,}027503(2012).

\bibitem{r1} Z. Roupas, Classical Quantum Gravity {\bf 30}, 115018 (2013).
\bibitem{r21} L. M. Cao, J. Xu, and Z. Zeng, Phys. Rev. D. {\bf 87}, 064005 (2013).
\bibitem{r22} L. M. Cao and J. Xu, Phys. Rev. D. {\bf 91}, 044029 (2015).
\bibitem{r3} Z. Roupas,  arXiv:1305.4851.
\bibitem{r4} N. Savvidou and C. Anastopoulos, Classical Quantum Gravity {\bf 31}, 055003 (2014).
\bibitem{r5} J. S. Schiffrin, arXiv:1506.00002.
\bibitem{r6} S. R. Green, J. S. Schiffrin and R. M. Wald, Classical Quantum Gravity {\bf 31} 035023 (2014).
\bibitem{r7} N.Savvidou and C.Anastopoulos, Classical Quantum Gravity {\bf 31}, 055003 (2014).

\bibitem{r8} R. Yang, Entropy 16, no. 8, 4483 (2014).

\bibitem{fang} X. Fang and S. Gao, Phys.Rev.D {\bf 90}, 044013(2014).

\bibitem{waldbook} R.M. Wald, {\em General Relativity} (University of Chicago, Chicago, 1984).


\end{thebibliography}
\end{document}